\documentclass[sigconf]{acmart}
\usepackage{amsmath,amsfonts}
\usepackage{algorithm}
\usepackage{algpseudocode}
\usepackage{graphicx}
\usepackage{textcomp}
\usepackage{xcolor}
\usepackage{soul}
\usepackage{subcaption}
\usepackage{enumitem}
\usepackage{listings}
\usepackage{float}
\usepackage{booktabs,tabularx,array}
\newcolumntype{Y}{>{\raggedright\arraybackslash}X}
\usepackage{comment}

\title{A Proof of Success and Reward Distribution Protocol for Multi-bridge Architecture in Cross-chain Communication}

\author{Damilare Peter Oyinloye}
\affiliation{%
  \institution{Department of Computer Science, NTNU}
  \city{Trondheim}
  \country{Norway}
}
\email{peter.d.oyinloye@ntnu.no}

\author{Mohd Sameen Chishti}
\affiliation{%
  \institution{Department of Computer Science, NTNU}
  \city{Trondheim}
  \country{Norway}
}
\email{mohd.s.chishti@ntnu.no}

\author{Jingyue Li}
\affiliation{%
  \institution{Department of Computer Science, NTNU}
  \city{Trondheim}
  \country{Norway}
}
\email{jingyue.li@ntnu.no}

\begin{document}

\title{A Proof of Success and Reward Distribution Protocol for Multi-bridge Architecture in Cross-chain Communication \\
}

\begin{abstract}

Single-bridge blockchain solutions enable cross-chain communication. However, they are associated with centralization and single-point-of-failure risks.
This paper proposes Proof of Success and Reward Distribution (PSCRD), a novel multi-bridge response coordination and incentive distribution protocol designed to address the challenges. 
PSCRD introduces a fair reward distribution system that equitably distributes the transfer fee among participating bridges, incentivizing honest behavior and sustained commitment. The purpose is to encourage bridge participation for higher decentralization and lower single-point-of-failure risks. 
The mathematical analysis and simulation results validate the effectiveness of PSCRD using two key metrics: the Gini index, which demonstrates a progressive improvement in the fairness of the reward distribution as new bridge groups joined the network; and the Nakamoto coefficient, which shows a significant improvement in decentralization over time. 
These findings highlight that PSCRD provides a more resilient and secure cross-chain bridge system without substantially increasing user costs.

\end{abstract}

\keywords{blockchain, multi-bridge, cross-chain, reward}

\maketitle

\section{Introduction} \label{sec:problem}

\begin{figure}
    \centering
    \includegraphics[width=0.48\textwidth]{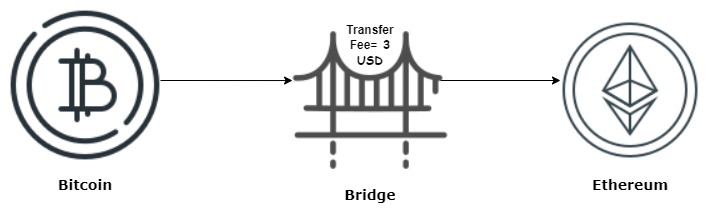}
    \caption{A Single Bridge Architecture}
    \label{fig:singlebridge}
\end{figure}

As blockchain adoption expands across finance, supply chain, healthcare and energy, it has produced various platforms with distinct consensus, data, and security models \cite{Inter1}. While this diversity supports sector-specific needs, it also creates significant interoperability challenges. Most blockchains remain isolated, restricting assets and data to their native environments \cite{literature4}, which obstructs seamless value transfer and information sharing \cite{Probstatement}.
Bridges are the most prominent interoperability solution \cite{Existingsolutions, Existingsolutions2}, enabling cross-chain functionality by monitoring and validating source-chain transactions and then triggering the corresponding operations on a destination chain \cite{Singlextra}. As shown in \autoref{fig:singlebridge}, this single-bridge model allows interoperability but introduces centralization risks and a single point of failure \cite{Bridgeproblem, bridgeproblem2}. 
To overcome this, researchers have proposed multi-bridge architectures, where multiple bridges connect two blockchains simultaneously to improve fault tolerance and decentralization \cite{uniswap, Multibridge1}. This approach mirrors safety strategies in critical systems such as aviation, where redundancy is a fundamental design principle \cite{aircraft}. Commercial aircraft, for instance, employ dual or multiple engines and redundant control systems to ensure continued operation even if one component fails \cite{aircraft2}. Similarly, multi-bridge systems aim to eliminate single points of failure in blockchain interoperability, thereby enhancing both reliability and systemic resilience.

While the multi-bridge concept offers architectural redundancy and resilience, it remains largely at the conceptual stage. The fundamental challenge extends beyond technical coordination to the economic dynamics among bridges. In practice, these entities are independent, profit-driven operators whose cooperation cannot be assumed. Without appropriate incentive mechanisms, bridges may lack motivation to participate, withhold responses, or even behave dishonestly. Consequently, the core research problem is to design a reward mechanism to ensure secure and economically sustainable cooperation among multiple bridges.

This paper proposes \textbf{Proof of Success and Reward Distribution~(PSCRD)}, a multi-bridge protocol centered on fair and transparent incentive allocation.
While PSCRD includes a coordination layer that organizes bridge responses and enables consensus on the destination blockchain, its core contribution lies in the reward mechanism.
By fairly distributing transfer fees among participating bridges, PSCRD promotes commitment, deters downtime or malicious activity, and enhances the security and robustness of cross-chain operations~\cite{Reward1}.

To evaluate PSCRD, we conducted simulations using the Gini index and the Nakamoto coefficient to measure fairness and decentralization. Results show improved equity in reward distribution as more bridge groups join the network, alongside a notable increase in decentralization.
These findings demonstrate PSCRD’s ability to align robustness with economic incentives, enabling resilient and fair cross-chain interoperability without imposing significant costs on end users.

The rest of this paper is organized as follows: Section~\ref{Sec:Literature} presents a comprehensive review of the relevant literature. Section~\ref{sec:design} introduces the proposed protocol. Section~\ref{sec:Proof} presents a mathematical proof to validate the correctness and fairness of the protocol's reward mechanism. Section~\ref{sec:Simulation} describes the evaluation using simulation and the results, while Section~\ref{sec:Discussion} discusses the results. Finally, Section~\ref{sec:Conclusion} concludes the paper.

\section{Literature Review} \label{Sec:Literature}

This review surveys state-of-the-art bridge architectures for cross-chain transfers, drawing from academic publications and industry white papers.
Zero-knowledge and cryptography-based approaches, such as zkBridge~\cite{zkbridge} and Axelar~\cite{axelar} emphasize strong security guarantees through cryptographic verification. zkBridge uses zero-knowledge proofs for trustless cross-chain verification, avoiding intermediaries but facing computational overhead and validator-entry barriers. Similarly, Axelar employs proof-of-stake validators, threshold cryptography, and light clients to validate cross-chain states, offering general message passing but also encountering risks of validator centralization and high maintenance costs. Both systems highlight the trade-off between robust cryptographic trust and efficiency.

Service- and notary-based architectures provide alternative designs. BlockChain I/O~\cite{blockchainIO} implements cross-chain services (CC-SVCs) with a modular PieChain core, reputation-based governance, and decentralized identity integration. While flexible, it risks centralization in the service layer and struggles with coordination across heterogeneous chains. Complementing this, an RNN-based method~\cite{literature3} enhances notary systems by introducing dynamic reputation decay and adaptive evaluation windows, improving detection of malicious actors but suffering from computational complexity and limited applicability to ecosystems without notaries. Together, these works demonstrate the tension between security monitoring and operational scalability.

Validator- and staking-driven solutions like Wanchain~\cite{wanchain2024} and Multichain~\cite{multichain2022} leverage multi-party computation (MPC) and secure multiparty computation (SMPC) respectively. Wanchain employs Storeman nodes, lock-mint-burn-unlock transactions, and staking-linked rewards, but faces scalability and stake allocation challenges. Multichain secures transfers with validator-generated threshold signatures and chain-specific adapters, though centralization risks, key management complexity, and smart contract vulnerabilities remain. Both illustrate how staking-based incentives can secure bridges but may introduce fragility under scaling pressure.

Liquidity- and network-based protocols such as Celer cBridge~\cite{celer2018} and THORChain~\cite{thorchain2020} focus on multichain interoperability and native asset swaps. Celer combines pool-based and canonical token bridges with coordination via the State Guardian Network, offering low-cost transfers but relying on intermediary trust and token value. THORChain, built as a Layer 1 DEX, enables continuous liquidity pools and native swaps secured by validator rotation and threshold signatures, but its complex validator system and transition dynamics risk instability. These models highlight the promise of liquidity-driven architectures but reveal challenges in ensuring resilience across chains.

\autoref{tab:bridge_comparison} presents a comparison of these solutions with PSCRD. These systems introduce innovations ranging from zero-knowledge proofs to MPC, staking, and liquidity pools; yet they share common drawbacks such as centralization, incentive complexity, and systemic fragility resulting from reliance on single-bridge mechanisms. These limitations underscore the need for more resilient and decentralized multi-bridge cross-chain solutions.

\begin{table*}[!t]
\centering
\scriptsize
\setlength{\tabcolsep}{3pt}
\caption{Comparison of bridges using fairness, cost and latency as metrics}
\label{tab:bridge_comparison}
\begin{tabularx}{\textwidth}{l Y Y Y Y Y Y Y Y Y}
\toprule
 & zkBridge & Axelar & BlockChain I/O & RNN-Notary & Wanchain & Multichain & cBridge & THORChain & \textbf{PSCRD(new)} \\
\midrule
\textbf{Fairness} &
No explicit fairness mechanism &
No explicit fairness mechanism &
No formal fairness metric &
No formal fairness metric &
No formal fairness metric &
No formal fairness metric &
liquidity-dependent &
No formal fairness metric &
\textbf{Clear fairness mechanism} \\
\addlinespace[2pt]
\textbf{Latency} &
High &
Moderate &
High &
High &
Moderate &
Moderate &
Low &
Low &
\textbf{Low (bridges runs in parallel)}\\
\addlinespace[2pt]
\textbf{Cost} &
Payment for a single bridge &
Payment for a single bridge &
Payment for a single bridge &
Payment for a single bridge &
Payment for a single bridge &
Payment for a single bridges &
Payment for a single bridge &
Payment for a single bridge &
\textbf{One payment for multiple bridges} \\
\bottomrule
\end{tabularx}
\end{table*}

\section{PSCRD Protocol} \label{sec:design}

\begin{figure}
    \centering
    \includegraphics[scale=0.30]{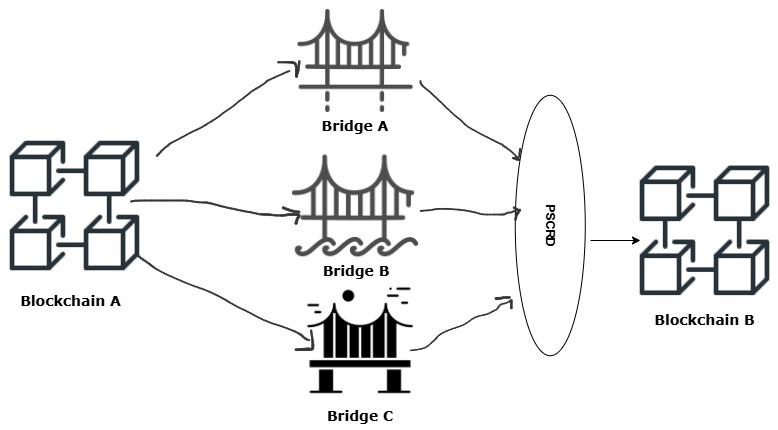}
    \caption{A Multi-Bridge Architecture with PSCRD}
    \label{fig:multibridgepos}
\end{figure}

To address the single point of failure and centralization challenges, our Proof of Success and Reward Distribution (PSCRD) protocol aims to coordinate multiple bridges for cross-chain communication in parallel and distribute awards fairly among the bridges.  
Our mechanism allows seamless transfers of data or assets between two distinct blockchain networks through multiple independent bridges. 
Each bridge facilitates the same transaction between the origin chain \texttt{Blockchain A} and the destination chain \texttt{Blockchain B}.
And these bridges are independent of each other and act as redundant pathways for the same transaction. ~\autoref{fig:multibridgepos} depicts this scenario with PSCRD as the coordination and reward distribution protocol.
For example, I want to transfer a token from Ethereum \texttt{Blockchain A} to Polygon \texttt{Blockchain B}. Using the multi-bridge architecture, I will initiate the transfer on Ethereum, and the transaction request is sent to 3 independent bridges. Each of these bridges attempts to transfer the token to Polygon. PSCRD will receive the bridges' responses and transmit the response agreed by the majority of the participating bridges, i.e., the majority response, to Polygon. The token is then credited to Polygon, which completes the process.

As multiple bridges are used simultaneously, a mechanism is required to coordinate their responses to ensure communication results correctness, fair incentivization, and security through decentralization. 
PSCRD coordinates the response of bridges and ensures that a consensus is reached at the destination blockchain, and each involved bridge is rewarded fairly based on its contributions, and the multi-bridge communication is decentralized to ensure its security. 

The PSCRD protocol has two essential components. The first is to select the participating bridges to form a quorum, and to achieve consensus on the responses on the destination chain. The underlying principle applied is majority voting. The second component computes rewards and applies reward decay to ensure equitable opportunities for all participating bridges. If the rewards are not distributed fairly, the possible consequence is that bridges are demotivated to participate, resulting in centralization and security risks. The protocol aims to achieve high reward fairness, measured using the Gini Index (see Section \ref{GiniIndex}), 
and high decentralization among the bridges, measured using Nakamoto Coefficient (see Section \ref{nakamotocoefficient}).

\subsection{Bridge selection and consensus achievement} 
To begin, a quorum $Q$ is computed as:  

\begin{equation}
Q = \frac{R_{total}}{R_{min}},
\end{equation}

where $R_{total}$ is the total reward to be distributed and $R_{min}$ is the minimum reward for each participating bridge. This ensures that the total reward pool is sufficient to pay all quorum bridges at the minimum reward rate. To satisfy the principle of impartial participation and ensure a fair opportunity for all bridges to be selected as part of $Q$, we randomly select $Q$ bridges given $N$ as the total number of bridges. The probability $P$ that any bridge $i$ will be selected as part of $Q$ is calculated as; 


\begin{equation}
P = \frac{Q}{N}.
\end{equation}

At the destination blockchain, the responses of the bridges will be grouped based on matching transaction details, such as the sender, receiver, and amount transferred. 
For a transaction to be considered valid on the destination blockchain, more than 50\% of the bridges (so-called the majority of bridges) must reach a consensus.

\subsection{
Reward distribution mechanism and decay conditions} \label{sec:rewards and penalty}

At the end of every transfer round, bridges belonging to the majority group are rewarded with success points $S_P$ for successfully passing a message from the source blockchain to the destination blockchain. 
The success point represents the cumulative history of successful transmissions by a participating bridge and determines the proportion of the reward it receives. This mechanism incentivizes bridges to consistently transmit correct and accurate data from the source to the destination blockchain.
After this, the transfer fee $x$ is distributed among the bridges in the majority group and added to the bridge reward $R$.

Each bridge $i$ in the majority group receives a reward based on its success points $S_{pi}$:

\begin{equation}
R_i = \left( \frac{S_{pi}}{\sum_{j \in \text{MAJORITY}} S_{pj}} \right) \cdot x,
\end{equation}

where $x$ is the transaction fee, $\sum_{j \in \text{MAJORITY}} S_{pj}$ is the total success points of the majority nodes.
This guarantees a proportional reward distribution among the majority group and incentivizes honest participation among them.
To ensure a balance in reward allocation and to reduce the monopoly of some bridges that might have accrued $S_p$ from dominating the entire reward architecture. A decay function is introduced, where time-sensitive regularization is utilized to limit the dominance of the early nodes. Let $A_i$ be the age (hours) of node $i$, $T_w$ be the decay time window, and $\lambda$ be the decay factor. If $A_i > T_w$, the decayed success point $S_{pi}'$ is computed as:

\begin{equation}
S_{pi}' = \frac{S_{pi}}{1 + \lambda A_i}.
\end{equation}

When $A_i = 0$, $S_{pi}' = S_{pi}$, this implies that there is no decay for new nodes. As $A_i$ increases, $S_{pi}'$ decreases, promoting decentralization.
The value of $\lambda$ ranges from 0 to 1, where values closer to 0 indicate gentle decay, and values closer to 1 indicate faster decay~\cite{decay1}.
Furthermore, in scenarios where a bridge temporarily exits the network and later rejoins, regardless of how frequently this occurs, such behavior does not adversely impact the network's overall performance or integrity. 
This is because the system is designed to retain each bridge's historical data, including its age and accumulated success points, for a long time (e.g., one year). 
These values are preserved during the bridge's absence and reinstated upon its return, ensuring continuity in its reputation history.
Consequently, the network remains robust against intermittent bridge participation, and the rejoining bridge resumes operations with its prior status fully intact. 
However, if a bridge is offline for more than one year, its history will be archived and will not be used in the reward computation. Algorithm \autoref{alg:success} presents the  pseudocode of the proposed protocol.

\begin{algorithm}
\caption{PSCRD Protocol}
\label{alg:success}
\begin{algorithmic}[1]
\State \textbf{Step 1: Define Constants and Variables}
\State Define \texttt{NUMBER OF NODES} as $N$, \texttt{MIN\_REWARD} as $R_{min}$, \texttt{TOTAL\_REWARD} as $R_{total}$
\State Define $Q$ as \texttt{QUORUM} 
\State Define $x$ as the transfer fee in USD
\State Define \texttt{TIME\_WINDOW} $T_w$ for reward decay evaluation
\State Define \texttt{DECAY\_FACTOR} $\lambda$
\State Create \texttt{SUCCESS\_POINTS[]} $S_p$ array for each node
\Statex \hrulefill

\State \textbf{Step 2: Quorum Selection}
\State Create function \texttt{Select\_Bridges($N$, $Q$)}
\State Compute $Q$ = $R_{total}$ / $R_{min}$
\State Randomly select $Q$ nodes from $N$
\State Return selected nodes as \texttt{Selected\_bridges}
\Statex \hrulefill

\State \textbf{Step 3: Transaction Processing and Response Grouping}
\State Create \texttt{Process\_Trans.(Quorum\_Bridges, Source\_Data)}
\State Initialize \texttt{RESPONSES} as an empty dictionary
\For{each bridge in \texttt{Quorum\_Bridges}}
    \State Request bridge to process \texttt{Source\_Data}
    \If{response exists in \texttt{RESPONSES}}
        \State Append bridge to response group
    \Else
        \State Create new entry and add bridge
    \EndIf
\EndFor
\State Identify \texttt{MAJORITY\_RESPONSE};
\State Return \texttt {MAJORITY RESPONSE}
\Statex \hrulefill

\State \textbf{Step 4: Reward Distribution}
\State Create \texttt{Distribute\_Rewards(MAJORITY\_RESPONSE, $x$)}
\State Initialize $S_p$ as an empty dictionary
\For{each bridge in \texttt{MAJORITY\_RESPONSE}}
    \If{bridge belongs to \texttt{MAJORITY\_RESPONSE}}
        \State Increment bridge's $S_p$ by 1
    \EndIf
\EndFor
\State Compute \texttt{TOTAL\_MAJORITY\_SUCCESS\_POINTS} $S_{p}$[j]
\For{each bridge in \texttt{MAJORITY\_RESPONSE}}
    \State $R$[i] = \texttt{($S_p$[i] / $\sum_{j \in \text{MAJORITY}} S_p$[j]) * $x$}
    
    \State Update bridge's $R$
\EndFor
\Statex \hrulefill

\State \textbf{Step 5: Reward Decay Mechanism}
\State Create \texttt{Apply\_Decay\_Mechanism($S_p$, $T_w$, $\lambda$)}
\For{each bridge in $Q$}
    \State Calculate $A$[i]
    \If{ $A_i > T_w$}
    $S_{pi}'$ = {${S_{pi}} / 1 + \lambda A_i$}
    \EndIf
\EndFor
\end{algorithmic}
\end{algorithm}

\subsection{Fairness and decentralization metrics}

The reward distribution and decay mechanism aim to ensure reward fairness and decentralization of the bridges. Fairness is measured using the Gini index and mathematically proven using the Lorenz curve. Decentralization is measured using the Nakamoto coefficient.

\subsubsection{Gini Index}
\label{GiniIndex}

The Gini index is a widely used statistical measure of economic inequality. It is one of the most popular metrics for assessing income or wealth distribution and is often used to aid in making informed policy decisions. 
In the context of blockchain, the Gini index measures how evenly rewards are distributed among participating bridges \cite{gini2}.
Its coefficient ranges from 0 to 1, where 0 indicates perfect equality (everyone receives an equal share) and 1 indicates perfect inequality (only one recipient or group receives all the income). 
The Gini index remains an essential tool for economists, sociologists, and policy makers to quantify and compare inequality, forming the basis for interventions aimed at reducing economic disparities~\cite{gini}.

\subsubsection{Lorenz Curve}
\label{lorenzcurve}

The Lorenz curve is a fundamental graphical representation that is used to visualize the distribution of income or wealth within a population. 
It plots the cumulative share of total income against the cumulative share of population, when units are arranged in ascending order of income \cite{curve1}. 
It is a graphical tool that provides an intuitive visualization of inequality, where perfect equality is represented by a straight diagonal line (the line of perfect equality), and any deviation from this line indicates the presence of inequality \cite{curve2}. 
Besides visualizing the distribution of income or wealth within a population, the Lorenz curve coordinates can be used to reflect the Gini index.

\subsubsection{Nakamoto Coefficient}
\label{nakamotocoefficient}

It is a quantitative measure of decentralization in blockchain networks, defined as the minimum number of independent entities (nodes) required to disrupt or take control of the network. 
It is often used to determine how decentralized a network is by evaluating the distribution of control across network participants. 
Named after Bitcoin's pseudonymous creator Satoshi Nakamoto, the coefficient provides a numerical assessment of the resilience of a blockchain to centralization risks \cite{nakamoto}. 
A higher Nakamoto Coefficient indicates greater decentralization and network security, as more independent entities would need to collude to compromise the system. 
This metric has become increasingly important in evaluating blockchain networks' security properties, guiding investors in risk assessment, helping developers identify centralization points, and informing governance decisions to maintain or improve decentralization in distributed systems \cite{nakamoto2}.

\section{Mathematical Proof} \label{sec:Proof}

\subsection{Defination}

Let $S_{pi}$ represent the success points accumulated by node $i$, for $i= 1,2,\ldots,n$. The decay-adjusted success points $S_{pi}'$ is given as;

    \begin{equation}
    S_{pi}' = \frac{S_{pi}}{1 + \lambda A_i},
    \end{equation}
    where $\lambda > 0$ is the decay factor and $A_i \geq 0$ is the age of the bridge. We have the following assumptions:
\begin{quote}
(\textbf{A1}) The decay is \textit{monotonic in success}, i.e., $S_{pi} > S_{pj} \Rightarrow A_i \geq A_j$. In other words, bridges with higher success points are also those with a higher age $A$, and are more significantly affected by decay.
\end{quote}
    
\subsection{Gini Index Formal Derivation}

Let the success points before and after decay be sorted from smallest to largest:
\[
S_{p(1)} \leq S_{p(2)} \leq \dots \leq S_{p(n)}, \quad
S'_{p(1)}\leq S'_{p(2)} \leq \dots \leq S'_{p(n)}.
\]
We define the total success points before decay $S$ and after decay $S'$  as
\begin{align}
    S &= \sum_{i=1}^n S_{pi} \quad \\
    S' &= \sum_{i=1}^n S_{pi}' \quad 
\end{align}

The Gini Index $G$ before decay and after decay is defined respectively:

\begin{equation}
    G = \frac{1}{n S} \sum_{i=1}^{n} (2i - n - 1) S_{pi}
\end{equation}
\begin{equation}
    G' = \frac{1}{n S'} \sum_{i=1}^{n} (2i - n - 1) S'_{pi},
\end{equation}

where $n$ is the total number of bridges, $S_{pi}$, $S'_{pi}$ are the sorted success points, $S$, $S'$ are the total of all success points before and after decay, respectively. 
The Gini index can be further defined in terms of the Lorenz curve, which maps the cumulative share of population (i.e., bridges) to the cumulative share of resources (i.e., success points). The Lorenz curve coordinates with respect to the population and resources are defined as;

\begin{equation}
L_k = \frac{1}{S} \sum_{i=1}^k S_{pi}, \quad
L'_k = \frac{1}{S'} \sum_{i=1}^k S'_{pi}
\end{equation}

Where $L_k$, $L'_k$ are the cumulative shares up to the $k^{th}$ bridge before and after decay, respectively. Furthermore, the Gini index before and after decay (i.e., $G_L$ and $G'_L$, respectively) can be expressed in terms of Lorenz curve coordinates:

\begin{equation}
G_L = 1 - \frac{2}{n} \sum_{k=1}^n L_k, \quad
G'_L = 1 - \frac{2}{n} \sum_{k=1}^n L'_k,
\end{equation}

Given that decay disproportionately affects bridges with higher success points - in agreement with Assumption (\textbf{A1}) - the resulting Lorenz curve after decay stochastically dominates its predecay counterpart.

\[
L'_k \geq L_k \quad \text{for all } k \in \{1, \ldots, n\}
\]

Hence:
\begin{equation}
\sum_{k=1}^n L'_k \geq \sum_{k=1}^n L_k \quad \Rightarrow \quad G'_L \leq G_L,
\end{equation}
which implies that if, after applying decay, the cumulative success share of the bottom $k$ bridges increases or remains unchanged at each level, then the overall inequality among all bridges has decreased or remained constant.
However, if the decay is not uniform (i.e., $A_i \neq A_j$ for some $i \neq j$), then:
\begin{equation}
\exists k: L'_k > L_k \quad \Rightarrow \quad G_L' < G_L.
\end{equation}

This implies that there exists at least one index $k$ such that after decay, the bottom $k$ nodes hold a greater share of the total than before decay. For example, after applying decay, even if it's a part of the population (e.g., the bottom 30\%) that gets relatively more reward share than before, then the overall inequality has strictly decreased. \textbf{Therefore, under monotonic decay, the mechanism strictly reduces inequality:}
\[
\boxed{G' < G}
\]

\subsection{Nakamoto Coefficient Formal Derivation}

Let the success points before and after decay sort in descending order:

\[
S_{p(1)} \geq S_{p(2)} \geq \cdots \geq S_{p(n)}
\]
\[
S'_{p(1)} \geq S'_{p(2)} \geq \cdots \geq S'_{p(n)}
\]

We define the cumulative sums of the success points of the top $k$ bridges before and after decay as follows:

\begin{align}
    C_k &= \sum_{i=1}^k S_{pi} \\
    C'_k &= \sum_{i=1}^k S'_{pi}
\end{align}

The Nakamoto Coefficient before and after decay (i.e., K and K') is given as:

\begin{align}
    K &= \min\left\{k : C_k \geq \frac{1}{2} S \right\} \\
    K' &= \min\left\{k : C'_k \geq \frac{1}{2} S' \right\},
\end{align}
where $K$ denotes the Nakamoto Coefficient and $k$ represents the running index (i.e., the position of the bridge in the sorted list).
We assume the same condition as stated above in (\textbf{A1}): It states that bridges with higher $S_{pi}$ are older and are more penalized, which compresses the head of the distribution. Hence:

\[
    C'_k < C_k \quad 
\]

This implies that more bridges are needed to accumulate $50\%$ of the total success points after decay $S'$ than before decay. Therefore, following the decay process, the Nakamoto Coefficient increases, indicating improved decentralization. Hence: 

\[
    K' > K
\]

In summary, the combined reward and decay mechanisms contribute to optimizing both fairness and decentralization within the system.

\section{Simulation} \label{sec:Simulation}
The objective of our simulation is to evaluate whether the rewards are fairly distributed among the bridges despite their different entry times and ages. 
To quantify the assessments, we employed the Gini index and the Nakamoto coefficient. 

%

\subsection{Simulation Setup}

The simulation was configured with a total of $50$ bridges that joined the network at different intervals to analyze how entry timing affects reward distribution. Bridges were divided into three groups: Group~1 with $20$ bridges at the start $(0h)$, Group~2 with $20$ bridges at $(40h)$, and Group~3 with $10$ bridges at $(60h)$. 
This staggered entry reflects how cross-chain bridge networks typically expand incrementally rather than all at once, as seen in systems like Wormhole, LayerZero, Axelar, and Cosmos \cite{wormhole2023docs},\cite {layerzero2022whitepaper},\cite {axelar2023network}.
Modeling incremental participation also serves as a robustness test, allowing us to examine whether the PSCRD protocol preserves fairness and decentralization when early entrants could otherwise dominate.

\subsection{Simulation Parameters}
\label{sec:parameter}
\begin{itemize}

\item Number of Bridges = $50$
\item Duration of Simulation = $150$ (Hours)
\item Decay Factor = $0.05$
\item Time Window = $5$ (Hours)
\item Success Points = Sampled from a Poisson distribution with mean value = $5$
\item Age of Bridges = Selected randomly from $0$ to $150$ (Hours)
\end{itemize}

We set the decay factor to 0.05, which falls within the typical range (0,1). This value offers a balanced trade-off between preserving historical reliability and promoting recent activity. It ensures that bridges are fairly rewarded for consistent participation, while allowing the system to remain adaptive and decentralized \cite{decay2}\cite{decay1}.
Success points were initialized using a Poisson distribution with a mean value of 5 to model natural variability in event occurrences. This reflects a moderate expected success rate, introducing realistic diversity between bridges at initialization.
The number of bridges was selected to ensure sufficient network diversity. 
A simulation duration of 150 hours was chosen to capture long-term behavioral dynamics. 
A 5-hour time window was applied to balance decay sensitivity and responsiveness. 
Bridge ages were randomized between 0 and 150 hours to reflect varied participation patterns and to evaluate the robustness of the decay mechanism  \cite{Poisson} \cite{randomBridgeAge}. 

\subsection{Simulation results} 

\begin{figure}
    \centering
    \includegraphics[width=0.50\textwidth]{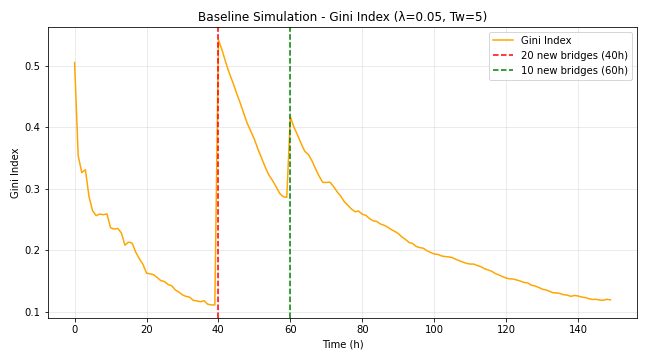}
    \caption{Gini Index for the Simulation}
    \label{fig:gini}
\end{figure}

\autoref{fig:gini} shows the system-wide Gini index throughout the 150-hour simulation, capturing how reward equity evolved as successive groups of bridges entered the network. 
The orange curve represents the Gini index values over time, serving as a measure of reward inequality among participating bridges. 
At the start, the simulation included 20 bridges (group 1). 
Two vertical reference lines mark the staggered introduction of additional groups: the red dashed line corresponds to group 2 (20 new bridges at 40h), while the green dashed line denotes group 3 (10 new bridges at 60h). 
Each new entry temporarily increased inequality, but the Gini index steadily declined afterward, converging to low values near 0.1, indicating that the rewards ultimately became equitably distributed across the 50 bridges.

\begin{figure}
    \centering
    \includegraphics[width=0.50\textwidth]{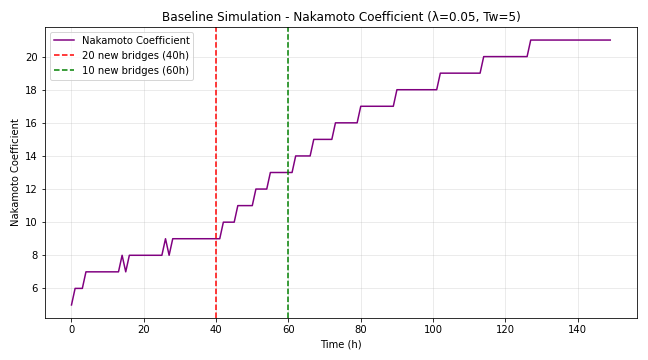}
    \caption{Nakamoto Coefficient for the Simulation}
    \label{fig:nakamoto}
\end{figure}

\autoref{fig:nakamoto} illustrates the evolution of the Nakamoto coefficient in the same period. 
The purple curve represents the coefficient over time, providing insight into the level of decentralization achieved in the system.
Initially, with only 20 bridges, the Nakamoto coefficient was relatively low. As additional groups joined, shown by the red and green dashed lines, the coefficient rose stepwise, stabilizing around 21 at the end of the simulation. 
This progression reflects how the inclusion of new bridges strengthened decentralization, demonstrating the ability of the protocol to maintain both equity and resilience as the network expanded.

\subsection{Parameter Sensitivity Analysis}
\label{sensitivityanalysis}

To evaluate the robustness of our parameter settings, we conducted a sensitivity analysis by varying the decay factor ($\lambda$) and the time window ($T_w$) used in the reward decay mechanism. Specifically, we tested $\lambda \in \{0.01, 0.05, 0.1\}$ and $T_w \in \{1, 5, 10\}$ hours, while keeping all other parameters identical to the baseline (50 bridges, simulation of 150 hours, success points sampled from a Poisson distribution with mean 5, and bridge groups joining at 0h, 40h, and 60h).

\begin{figure}
    \centering
    \includegraphics[width=0.50\textwidth]{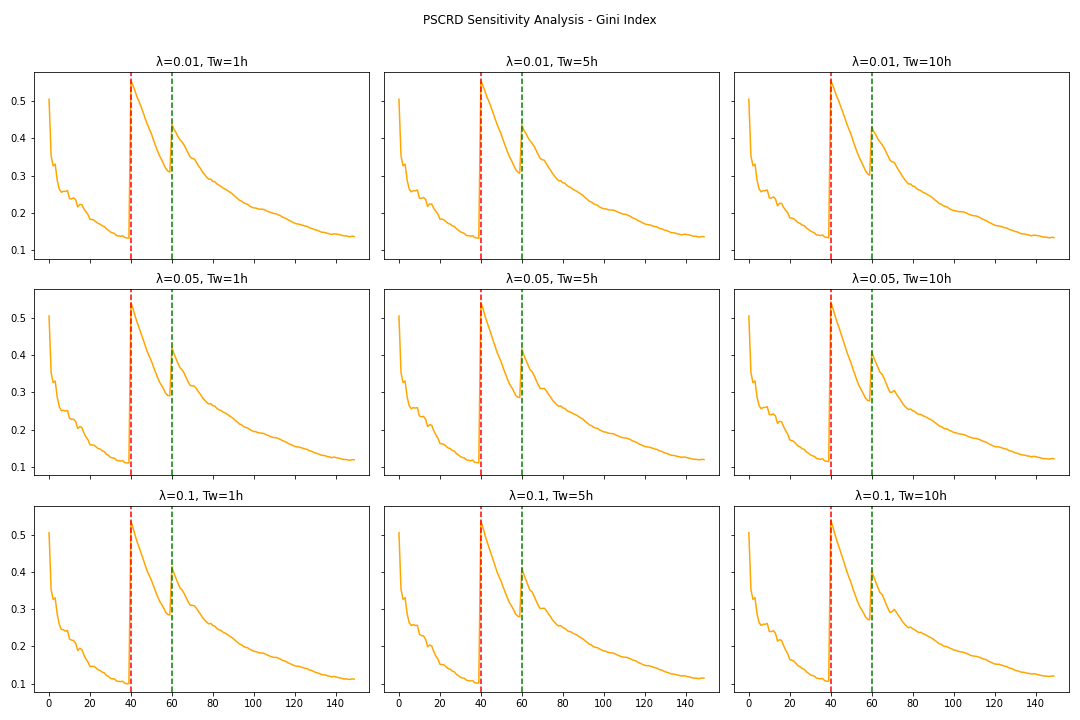}
    \caption{Gini Index Parameter Sensitivity}
    \label{fig:ginisensitivity}
\end{figure}

\begin{figure}
    \centering
    \includegraphics[width=0.50\textwidth]{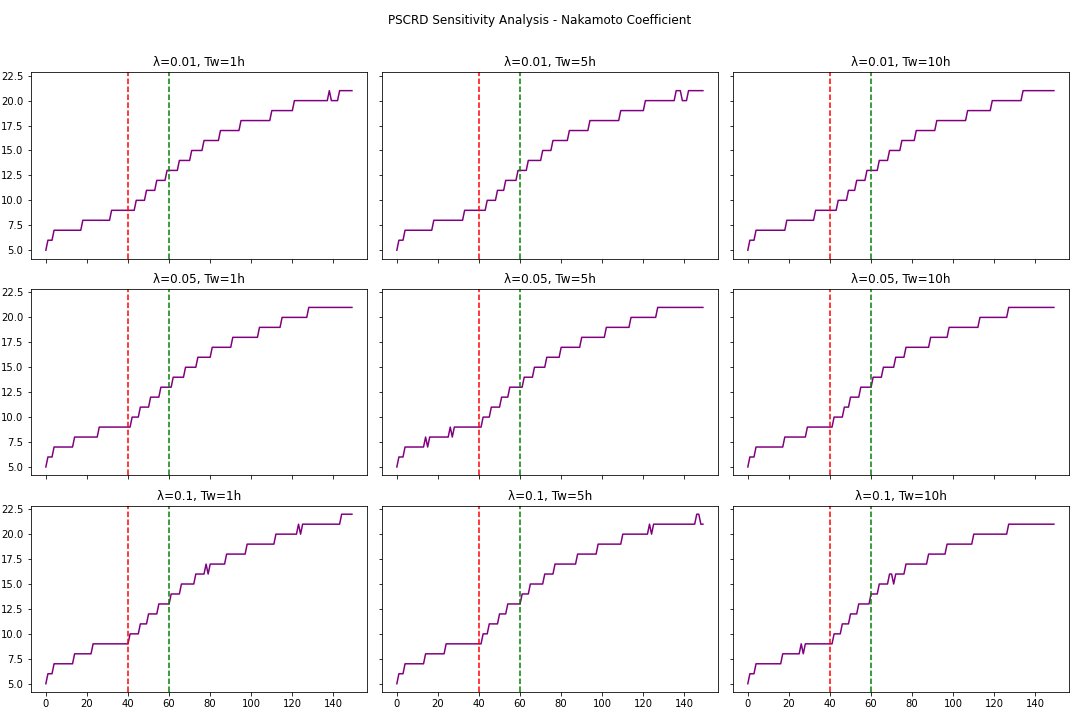}
    \caption{Nakamoto Coefficient Parameter Sensitivity}
    \label{fig:nakamotosensitivity}
\end{figure}

\autoref{fig:ginisensitivity} and ~\autoref{fig:nakamotosensitivity} show the resulting system-wide Gini index and Nakamoto coefficient over time. Across all parameter settings, the Gini index decreases steadily, converging to low values (0.1--0.15), which indicates that rewards remain fairly distributed. Similarly, the Nakamoto coefficient consistently increases as new bridge groups join, stabilizing around 20--22 and reflecting strong decentralization across all tested parameter choices. These results demonstrate that the PSCRD protocol is robust: while $\lambda$ and $T_w$ slightly affect the speed of convergence, the qualitative behavior is consistent. 

\subsection{Security Analysis} 
\label{sec:security}

The simulation was conducted with \(50\) bridges for \(150\) hours, while all other parameters remained the same as described in \autoref{sec:parameter}.
One critical security risk of multi-bridge cross-chains is the 51\% attack \cite{51Percentattack} . 
\textbf{PSCRD} mitigates the influence of 51\% attacks through two mechanisms. %
The first is \textbf{Randomized quorum selection}, which significantly reduces the likelihood of attacker dominance even when the attackers control 51\% of the participating bridges. The probability that attackers form a quorum majority in any given round is low and decreases rapidly with quorum size \(Q\), following the binomial tail.
The second mechanism is \textbf{Decay}, which removes any buildup of advantage even when the number of adversaries is 51\% or more.
Temporary control of the majority can raise  \( S_p \) short-term, but with the decay mechanism, they do not translate into sustained control or profit.  

We simulated the PSCRD protocol under two scenarios of a 51\% attack:  
(i) a baseline with 5 attackers, and  
(ii) a 51\% attack with 26 attackers out of the network’s 50 bridges.  
The results are presented in \autoref{tab:attacker_performance}.
In the baseline scenario, attackers rarely formed a quorum majority; therefore, their cumulative \( S_p \) and reward share remained approximately zero.  
In the scenario with 26 attackers, the average accumulated \(S_p\) of the attackers was \(2503.43\) without the decay function.  
When decay was applied, the cumulative \(S_p\) of the attackers decreased to \(279.39\), representing a substantial reduction. 
The reward share of the attackers at the end was \(49.73\%\) of the total system rewards, while the other bridges maintained the remaining \(50.27\%\), indicating no sustained economic advantage for the attackers.

\begin{table}
\centering
\caption{51\% Attack Analysis}
\begin{tabular}{lcc}
\toprule
\textbf{Metric} & \textbf{ Baseline-5 Attackers} & \textbf{26 Attackers} \\
\midrule
Attacker $S_p$ (-decay) & 0 & 2503.43 \\
Attacker $S_p$ (+ decay) & 0 & 271.39 \\
Attacker reward share & 0\% & 49.73\% \\
\bottomrule
\end{tabular}
\label{tab:attacker_performance}
\end{table}

\section{Discussion}
\label{sec:Discussion}

The evolution of the Gini index over the simulation period provides insight into the fairness of the reward distribution mechanism within the PSCRD architecture. 
In the initial phase of the simulation $(0h–40h)$, when only Group~1 (comprising $20$ bridges) was active, the Gini index exhibited a downward trend. 
Although this group held a substantial portion of the accumulated rewards due to their early, unchallenged participation, the decline in inequality indicates that the reward distribution was gradually becoming more equitable. 
At $40h$, the introduction of Group~2 ($20$ additional bridges) caused a temporary increase in inequality, as late participants began without prior rewards. However, this effect was transient: the index soon resumed its decline as new participants accumulated success points. 
A similar pattern was observed when Group~3 ($10$ bridges) joined at $60h$, producing a modest rise in inequality followed by a steady downward adjustment. 
By the end of the $150h$ simulation, the Gini index converged to a low value ($\approx 0.12$), confirming that the protocol promotes fairness over time even under staggered entry conditions.

The Nakamoto coefficient complements this analysis by quantifying the degree of decentralization in the network. 
During the initial phase $(0h–40h)$, the coefficient remained low, reflecting the dominance of Group~1. 
At $40h$, the entry of Group~2 bridges produced a noticeable upward shift, which was reinforced by the arrival of Group~3 at $60h$. 
As rewards became more widely distributed, the coefficient rose in a stepwise fashion and stabilized at around $21$ by the end of the simulation. 
This progression indicates that the PSCRD architecture naturally counterbalances early concentration of influence, as later entrants progressively dilute the dominance of initial participants.
To further evaluate the robustness of these findings, we conducted a sensitivity analysis by varying the decay factor ($\lambda$) and the time window ($T_w$) used in the reward decay mechanism (see Section \ref{sensitivityanalysis}).
%
Across all parameter combinations, the qualitative dynamics remained consistent: the Gini index decreased steadily toward values between $0.1$ and $0.15$, while the Nakamoto coefficient increased stepwise with new bridge arrivals, stabilizing between $20$ and $22$. 
These results demonstrate that although $\lambda$ and $T_w$ influence the speed of convergence, they do not alter the overall fairness or decentralization properties of the protocol. 
%
%
Taken together, these findings highlight the robustness of the PSCRD reward mechanism: it not only self-corrects initial imbalances as the network expands but also maintains fairness and decentralization under a wide range of parameter choices.

\section{Conclusion and Future Work} \label{sec:Conclusion}

In this paper, we propose PSCRD, a majority-voting-based coordination and reputation-based reward distribution protocol designed for a multi-bridge architecture in cross-chain communications. 
The protocol facilitates reliable coordination between the destination blockchain and the multiple bridges that transfer assets and data from the source blockchain.
It classifies bridge responses, identifies the majority group based on metadata, and allocates rewards proportionally based on each bridge’s earned success points.  
We mathematically analyzed the fairness and decentralization of the reward mechanism using the Gini index and the Nakamoto coefficient, and further examined its robustness through a security analysis demonstrating resistance to 51\% attacks and a parameter sensitivity analysis validating stability under varying network conditions. 
Simulation results show the effectiveness of the proposed protocol in enhancing fairness and decentralization within a multi-bridge cross-chain architecture.
%
%
Future work will focus on applying this architecture to relay DAO governance decisions from the main blockchain to remote deployments, thereby improving the security and extensibility of DAO governance mechanisms.

\bibliographystyle{ACM-Reference-Format}

\bibliography{Blockchain}

\end{document}